\documentclass[11pt,twoside]{article}

\usepackage{amsfonts}
\usepackage{amsmath}
\usepackage{amsthm}
\usepackage{amscd}
\usepackage{cite}
\usepackage{eucal}
\usepackage{graphicx}
\usepackage{a4}
\usepackage{microtype}
\usepackage{hyperref}
\usepackage{times}
\usepackage{float}

\def\beq{\begin{equation}}
\def\eeq{\end{equation}}
\def\bea{\begin{eqnarray}}
\def\eea{\end{eqnarray}}

\def\beann{\begin{eqnarray*}}
\def\eeann{\end{eqnarray*}}

\let\a=\alpha  \let\g=\gamma \let\de=\delta
\let\e=\varepsilon   

\let\dh=\vartheta  \let\la=\lambda 
\let\n=\nu  \let\p=\pi  \let\s=\sigma
 
\let\ph=\varphi   
  
  \let\D=\Delta

\let\qd=\quad  

\def\epp{\, .}
\def\epc{\, ,}

\theoremstyle{plain}

\newtheorem*{corollary*}{Corollary}

\newtheorem*{conjecture*}{Conjecture}

\theoremstyle{definition}

\def\2{\frac{1}{2}} \def\4{\frac{1}{4}}

\def\6{\partial}

\def\+{\dagger}

\def\<{\langle} \def\>{\rangle}

\def\i{{\rm i}}

\def\rd{{\rm d}}
\def\re{{\rm e}}

\DeclareMathOperator{\sh}{sh}

\DeclareMathOperator{\arctg}{arctg}
\DeclareMathOperator{\arcctg}{arcctg}

\DeclareMathOperator{\sn}{sn}

\DeclareMathOperator{\dn}{dn}

\def\Re{{\rm Re\,}}

\begin{document}

\thispagestyle{empty}

\begin{center}

{\Large {\bf Asymptotics of correlation functions of the Heisenberg-Ising
chain in the easy-axis regime \\}}

\vspace{7mm}

{\large
Maxime Dugave,\footnote{e-mail: dugave@uni-wuppertal.de}
Frank G\"{o}hmann\footnote{e-mail: goehmann@uni-wuppertal.de}}%
\\[1ex]
Fachbereich C -- Physik, Bergische Universit\"at Wuppertal,\\
42097 Wuppertal, Germany\\[2.5ex]
{\large Karol K. Kozlowski\footnote{e-mail: karol.kozlowski@u-bourgogne.fr}}%
\\[1ex]
IMB, UMR 5584 du CNRS,
Universit\'e de Bourgogne, France\\[2.5ex]
{\large Junji Suzuki\footnote{e-mail: sjsuzuk@ipc.shizuoka.ac.jp}}%
\\[1ex]
Department of Physics, Faculty of Science, Shizuoka University,\\
Ohya 836, Suruga, Shizuoka, Japan

\vspace{35mm}

{\large {\bf Abstract}}

\end{center}

\begin{list}{}{\addtolength{\rightmargin}{9mm}
               \addtolength{\topsep}{-5mm}}
\item
We analyze the long-time large-distance asymptotics of the
longitudinal correlation functions of the Heisenberg-Ising
chain in the easy-axis regime. We show that in this regime
the leading asymptotics of the dynamical two-point functions
is entirely determined by the two-spinon contribution to their
form factor expansion. Its explicit form is obtained from
a saddle-point analysis of the corresponding double integral.
It describes the propagation of a wave front with velocity
$v_{c_1}$ which is found to be the maximal possible group velocity.
Like in wave propagation in dispersive media the wave front
is preceded by a precursor running ahead with velocity $v_{c_2}$.
As a special case we obtain the explicit form of the asymptotics
of the auto-correlation function.
\end{list}

\clearpage

\section{Introduction}
New techniques for trapping and controlling cold atomic gases have
recently given us experimental access to the real-time dynamics
of correlated quantum systems. Dynamical correlation functions 
of atomic degrees of freedom in optical lattices can by now
be observed with single-site resolution \cite{BGPFG09,SWECBK10}.
Two-point functions describe the response to a local perturbation
spreading out into the system. So far relatively little is
known about these functions from the theory side. In this work
we calculate the asymptotic form of the dynamical two-point
functions of two local operators measuring the spin projection
onto the anisotropy axis in the Heisenberg-Ising chain in the
massive antiferromagnetic regime for long times and far
separated points. We may think of these functions as describing
the propagation of a signal through the quantum spin chain or
quantum communication. Their asymptotics describes the shape
of the signal at late times and far away from the source.

The Heisenberg-Ising chain is the spatially one-dimensional variant
of the fundamental model of antiferromagnetism in solids, the
anisotropic Heisenberg model. In nature it is realized in strongly
anisotropic solids which, since rather recently, can be simulated by
systems of trapped ions in optical lattices \cite{FukuharaSchauss,%
HildFukuhara}. The model is described by the Hamiltonian
\begin{equation} \label{ham}
     H = J \sum_{j = 1}^L \bigl\{ \s_{j-1}^x \s_j^x + \s_{j-1}^y \s_j^y
                               + \D \s_{j-1}^z \s_j^z \bigr\} \epc
\end{equation}
where $\s_j^\a$, $\a = x, y, z$, is a Pauli matrix acting on
site $j$ of a chain of length $L$, $J > 0$ quantifies
the exchange interaction, and $\D \in {\mathbb R}$ is
the anisotropy parameter. The model with $\D = 1$ is
the Heisenberg (or isotropic) model describing pure
antiferromagnetic exchange. If $\D \ne 1$ the Hamiltonian
is a linear combination of the Heisenberg and the Ising
chain Hamiltonians. $\D = 1$ separates two ground-state
regimes, the massless antiferromagnetic regime with
$|\D| < 1$ and the massive antiferromagnetic regime with
$\D > 1$ \cite{YaYa66d}.

The Heisenberg-Ising chain belongs to the class of
Yang-Baxter integrable models. Its ground state properties
were calculated long ago by means of the Bethe ansatz
\cite{Bethe31,Orbach58,YaYa66b,YaYa66c,YaYa66d}.
Subsequently, its excitation spectrum was analyzed
\cite{Woynarovich82c,BVV83,ViWo84} and a number of
methods to study its correlation functions were devised
(see e.g. \cite{JMMN92,JiMi96,KMT99b,GKS05,BJMST08a,%
JMS08,BoGo09,KKMST11b}). The cited articles clarify the
structure of the static correlation functions and provide
tools for their exact calculation at short and large distances.
Still, less is known about the dynamical correlation
functions of this well-studied model.

All results on dynamical correlation functions obtained
so far by exploiting the integrable structure of the model
rely on the calculation of matrix elements of local operators
in spectral representations of the correlation functions.
Matrix elements between ground state and excited states
are called form factors. In the thermodynamic limit
($L \rightarrow \infty$) they scale as $L^{- \de}$
with $\de \ge 0$. A finite number of them may stay finite
as $L \rightarrow \infty$, but all others must vanish for
the spectral representation to exist. Determinant formulae
for form factors of the Heisenberg-Ising chain at finite
$L$ were obtained in~\cite{KMT99a}. Their scaling behaviour
is different for different excitations and also differs
in the massless and in the massive regime. In the massless
regime $\de$ may be non-integer and is then called anomalous
dimension. The (anomalous) large-$L$ behaviour of so-called
particle-hole excitations for $|\D| < 1$ and a magnetic
field $h > 0$ in the direction of the anisotropy axis
was studied in \cite{KKMST09a,KKMST11a}. The summation
of the particle-hole form factors \cite{KKMST11b} resulted
in explicit formulae for the large-distance asymptotics
in the static case. In the dynamical case other types of
excitations (so-called strings) might contribute to the
asymptotics. For this reason their calculation is still
open.

For $\D > 1$ and $h$ below a certain critical field all
excitations over the ground state of the system in the
thermodynamic limit can be classified as scattering states 
of an even number of `spinons'. In this case the large
$L$ behaviour of \emph{all} form factors of the operator
$\s^z$ was recently obtained in \cite{DGKS15a}. The scaling
dimensions $\de$ are even integers, and the sums in the
spectral representation of two-point functions turn into
integrals over form factor densities at large $L$. Expressions
for these densities were first obtained in \cite{JMMN92}
within the $q$-vertex operator approach \cite{JiMi95}
which is a lattice version of bosonization in
two-dimensional quantum field theories.

The most successful application of form factor techniques
to the calculation of dynamical correlation functions so far
is in the approximate calculation of dynamical structure
factors. For $\D > 1$ results from the $q$-vertex operator
approach enabled the calculation of the two- and four-spinon
contributions to these quantities \cite{BKM98,HaCa06,CKSW12}.
The accuracy of these results was impressively demonstrated
in neutron scattering experiments (e.g.\ \cite{MourigalEnderle}).
Also the calculation of dynamical structure factors based
on a numerical evaluation of the finite-$L$ form factors
from Bethe ansatz proved to be efficient \cite{BKM03,%
SST04,CaMa05,PSCHMWA07}.

\section{Form-factor expansion}
In this work the focus is on explicit analytical results
for real-time correlation functions. The longitudinal
ground-state two-point functions of the Heisenberg-Ising
chain have the form-factor expansion (see \cite{DGKS15a})
\begin{multline} \label{twospexp}
     \bigl\< \s_1^z \s_{m+1}^z (t) \bigr\> \\ =
	\frac{(q^2; q^2)^4}{(-q^2; q^2)^4} (-1)^m
        + \sum_{\substack{\iota = 0, 1 \\ n \in 2 {\mathbb N}}} \frac{(-1)^{\iota m}}{n!}
	  \int_{- \frac{\p}{2}}^\frac{\p}{2} \frac{\rd^n \n}{(2 \p)^n}
	       {\cal F}_{\iota, n}^{(z)}
	       \prod_{j=1}^n \re^{\i [p(\n_j) m - \e (\n_j) t]}.
\end{multline}
Here we used the standard notation for $q$-multi factorials,
\begin{equation}
     (a;q_1, \dots, q_p) =
        \prod_{n_1, \dots, n_p = 0}^\infty (1 - a q_1^{n_1} \dots q_p^{n_p}) \epp
\end{equation}
The parameter $q = \re^{- \g} \in (0,1)$ in (\ref{twospexp}) is related
to the anisotropy parameter, $\D = (q + q^{-1})/2$. The time-independent
first term on the right hand side of (\ref{twospexp}) signifies the
antiferromagnetic long-range order for $\D > 1$. The dynamical information
is in the second term. The summation over $\iota$ is due to the double
degeneracy of the ground-state, while $n$ runs over all pairs of spinons.
With $p$ and $\e$ we have denoted momentum and energy of a single spinon.
They can be expressed as functions of a rapidity variable $\n$,
\begin{subequations}
\begin{align}
     p(\n) & = \frac{\p}{2} + \n
          - \i \ln \biggl( \frac{\dh_4 (\n + \i \g/2, q^2)}
	                        {\dh_4 (\n - \i \g/2, q^2)} \biggr) \epc \\[1ex]
     \e(\n) & = - \frac{4 J K \sh (\g)}{\p}
                 \dn \biggl( \frac{2 K \n}{\p} \bigg| k \biggr) \epp
\end{align}
\end{subequations}
Here $\dh_4$ is a Jacobi theta function and $\dn$ a
Jacobi-elliptic function, $k = k(q)$ is the elliptic modulus,
and $K = K(k)$ the complete elliptic integral of the first kind
(see e.g.\ \cite{WhWa63ch2122}). Spinon energy and momentum are
related by
\begin{equation} \label{pp}
     p' (\n) = - \e (\n)/2 J \sh(\g)
\end{equation}
and by the dispersion relation
\begin{equation} \label{disp}
     \e (p) = - \sqrt{v_{c_1} v_{c_2}} \cdot \sqrt{1/k^2 - \cos^2 (p)}
\end{equation}
which reveals the massive nature of the excitations. Here we have
introduced two combinations of parameters, which will become
important below,
\begin{equation}
     v_{c_1} = \frac{4 J K k^2 \sh(\g)}{\p (1 + k')} \epc \qd
     v_{c_2} = \frac{4 J K k^2 \sh(\g)}{\p (1 - k')} \epc
\end{equation}
where $k' = \sqrt{1 - k^2}$ is the complementary modulus.

The form factor densities ${\cal F}_{\iota, n}^{(z)}$ depend on
an even number $n$ of rapidities $\n_1, \dots, \n_n$. For general
$n$ they were expressed by multiple integrals in \cite{JiMi95}
and by Fredholm determinants in \cite{DGKS15a}. A more explicit
expression is only known for the two-spinon case $n = 2$ and was
extracted in \cite{DGKS15a} from Lashkevich's result \cite{Lashkevich02}
for the form factors of the XYZ chain in the corresponding limit.
It allows us to write the two-spinon contribution to the longitudinal
correlation function as
\begin{equation} \label{twospinon}
     \bigl\< \s_1^z \s_{m+1}^z (t) \bigr\>_2 =
	\frac{(q^2; q^2)^4}{(-q^2; q^2)^4} (-1)^m
	+ \2 I_2 (m, t) \epc
\end{equation}
where
\begin{equation} \label{imt}
     I_2 (m,t) = \biggl[ \prod_{j=1}^2
              \int_{- \frac{\p}{2}}^\frac{\p}{2} \frac{\rd \n_j}{2 \p}
	      \re^{\i [p(\n_j) m - \e (\n_j) t]} \biggr] f(\n_1, \n_2) \epp
\end{equation}
The function $f$ can be expressed as
\begin{equation}
     f(\n_1, \n_2) = A(\n_1, \n_2) (-1)^m + A(\n_1 + \p, \n_2)
\end{equation}
where
\begin{multline} \label{ampfun}
     A(\n_1, \n_2) = \frac{32 q (q^2;q^2)^2 \cos^2 \bigl( (p(\n_1) + p(\n_2))/2 \bigr)
	\sin^2 (\n_{12}) \dh_4^2 \bigl( \n_{12}/2, q \bigr)}
	     {\sin ((\n_{12} + \i \g)/2) \sin ((\n_{12} - \i \g)/2)} \\[1ex]
	\times \prod_{\s = \pm}
        \frac{(q^4;q^4,q^4)^2}{(q^2;q^4,q^4)^2}
        \frac{(q^4 \re^{2 \i \s \n_{12}};q^4,q^4)^2}
	     {(q^2 \re^{2 \i \s \n_{12}};q^4,q^4)^2}
	\frac{(q^4 \re^{2 \i \s \n_{12}};q^4)}
	     {(q^2 \re^{2 \i \s \n_{12}};q^4)}
\end{multline}
and $\n_{12} = \n_1 - \n_2$.
\section{Saddle point analysis}
\begin{figure}
\begin{center}
\begin{tabular}{ccc}
\includegraphics[width=.32\textwidth]{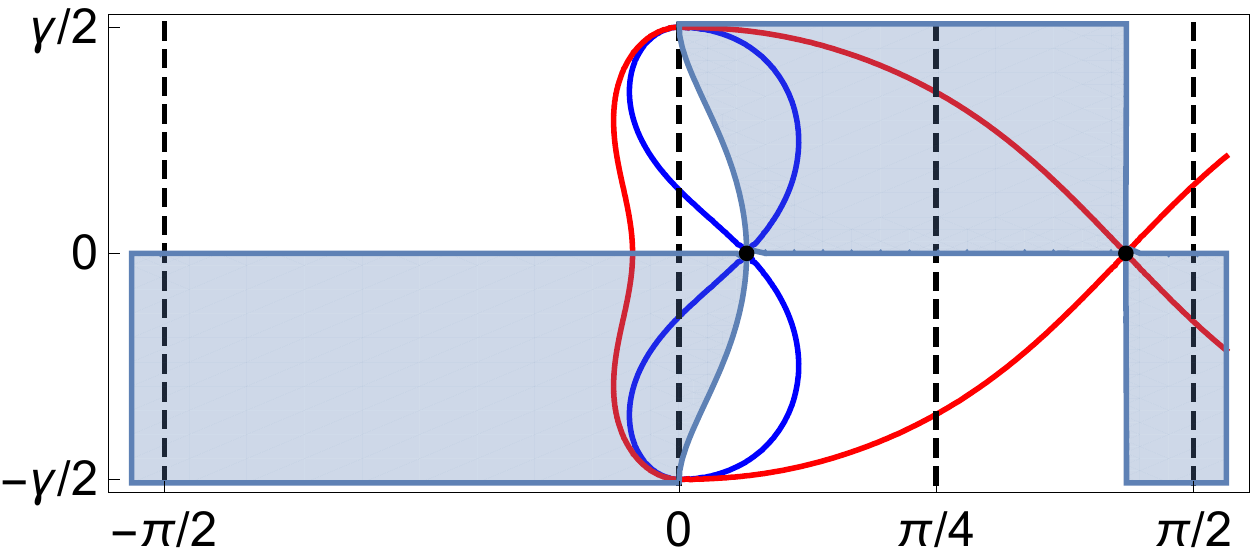}
\includegraphics[width=.32\textwidth]{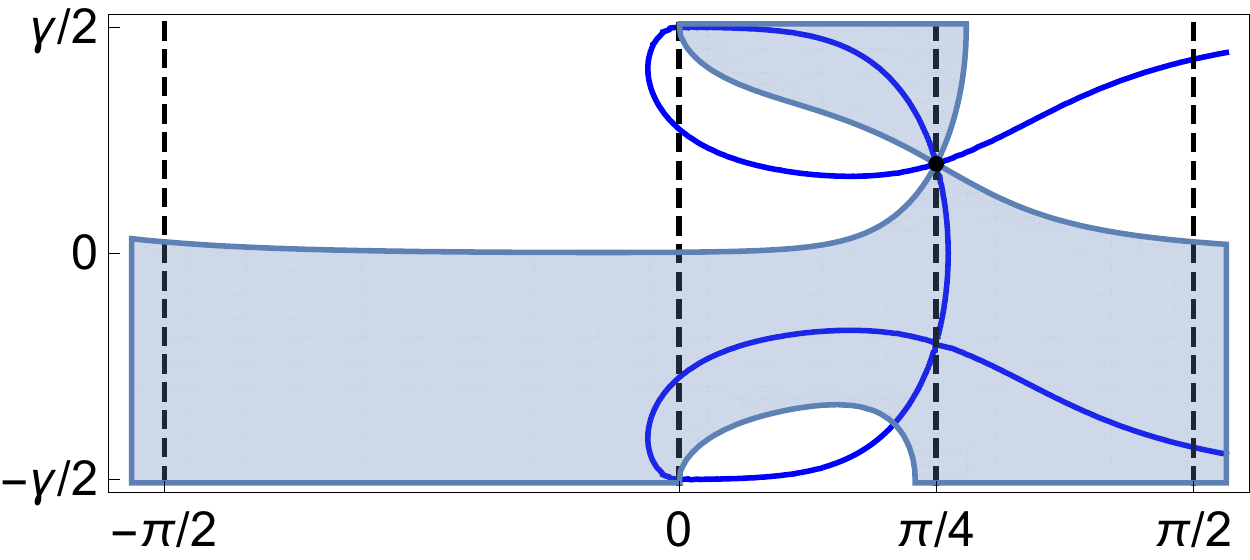}
\includegraphics[width=.32\textwidth]{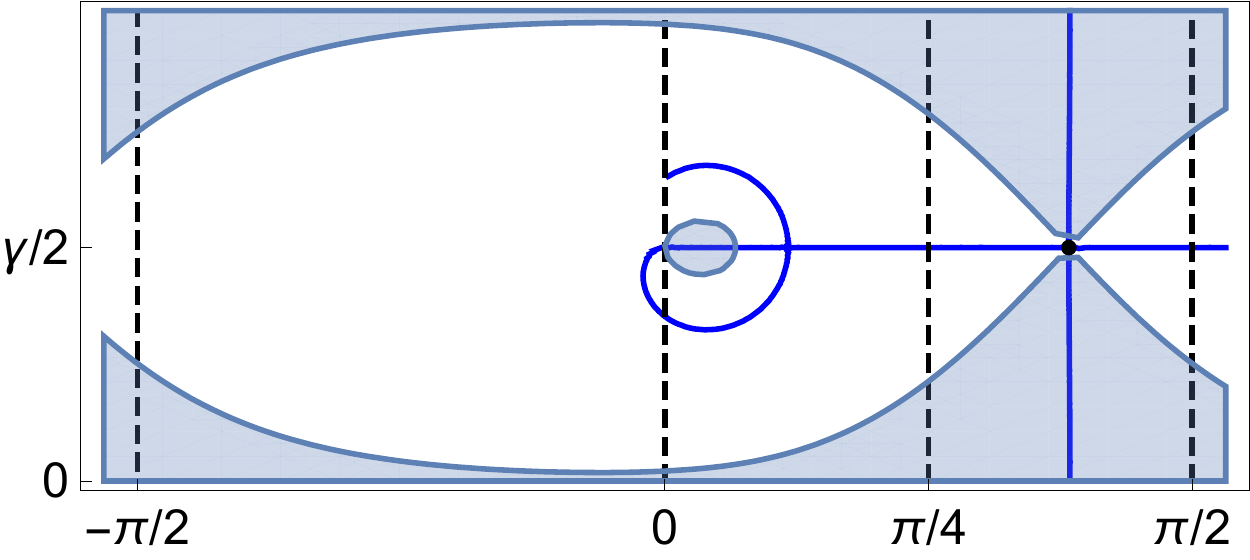}
\end{tabular}
\caption{\label{fig:phase} Behaviour of the phase $g$ in the
complex plane in the asymptotic regimes R1 - R3 (left to right).
The black dots denote the relevant saddle points, and the
lines are lines of constant imaginary part. In the
white areas $\Re g$ is smaller than the saddle point
values, in the blue shaded area it is lager. Contours
must be deformed in such a way that they pass through the
relevant saddle points and do not leave the white areas.}
\end{center}
\end{figure}

We have calculated the asymptotics $m, t \rightarrow \infty$
for fixed ratio $v = m/t \ge 0$. In this limit the integral $I_2$
can be estimated by the method of steepest descent. The calculation
shows that the higher-spinon contributions neglected in (\ref{twospinon})
contribute only higher corrections to the steepest descent result.
Let us define the function $g(\la) = \i [p(\la) v - \e (\la)]$.
Then the phase in the integrand in (\ref{imt}) is $g(\la) t$. It
is depicted in Figure~\ref{fig:phase}. The asymptotics of the
integral $I_2$ is determined by the roots of the saddle point
equation $g' (\la) = 0$ on steepest descent contours joining $-\p/2$
and $\p/2$. Define $k_1 = v_{c_1}/v_{c_2} = (1 - k')/(1 + k')$,
$K_1 = K(k_1)$. Then, using a Landen transformation, we can
write the saddle point equation as
\begin{equation} \label{spnu}
     \sn \bigl( 4 K_1 \la/\p \big| k_1 \bigr) = v/v_{c_1} \epp
\end{equation}
Its solutions divide the `$m$-$t$ world plane' into three different
asymptotic regimes R1, R2 and R3 .
\begin{itemize}
\item[R1.]
The `time-like regime' $0 < v < v_{c_1}$: In this case (\ref{spnu})
has two real solutions $\la_1^- < \la_1^+$ in $[-\p/2, \p/2]$ which
are located in $[0, \p/2]$ such that $\la_1^+ = \p/2 - \la_1^-$.
\item[R2.]
The `precursor regime' $v_{c_1} < v < v_{c_2}$: In this case (\ref{spnu})
has no real solutions. Setting $\la_2 = \p/4 + \i y$ it turns into
\begin{equation} \label{sp2nu}
     \dn \bigl( 2 K_1' y/\g \big| k_1' \bigr) = v_{c_1}/v
\end{equation}
which has real solutions $\pm y \in [- \g/2, \g/2]$ as long as
$v_{c_1} < v < v_{c_2}$ (here $k_1' = \sqrt{1 - k_1^2}$ and $K_1'
= K(k_1')$).
\item[R3.]
The `space-like regime' $v_{c_2} < v$: In this regime we introduce
$\la_3 = \i \g/2 + x$. Then (\ref{spnu}) implies
\begin{equation} \label{sp3nu}
     \sn \bigl( 4 K_1 x/\p \big| k_1 \bigr) = v_{c_2}/v
\end{equation}
which has again two solutions $x_- < x_+$ in $[-\p/2, \p/2]$,
both located in $[0, \p/2]$, such that $x_+ = \p/2 - x_-$.
\end{itemize}
Equations (\ref{spnu})-(\ref{sp3nu}) can be inverted as incomplete
elliptic integrals to give $\la_1^\pm$, $x$ and $y$ as functions
of $v$. Using these values we obtain the leading large-$t$ asymptotics
of $I_2$.

In R1 we find
\begin{equation} \label{i2r1}
     I_2 (m, t) \sim \frac{f(\la_1^+, \la_1^-)}{\p t}
                     \prod_{\s = \pm} \frac{\re^{t g(\la_1^\s)}}
		                           {\sqrt{g''(\la_1^\s)}} \epp
\end{equation}
Since $g(\la_1^\pm)$ is purely imaginary $I_2$ shows oscillations
and algebraic decay in R1. Note that we have obtained a factor of
$1/\sqrt{t}$ per integration. Assuming regular behaviour of the
four-spinon density ${\cal F}_{\iota, 4}^{(z)}$ one would at least
obtain a factor of $1/t^2$ which is sub-leading in R1 (in fact,
from our result in \cite{DGKS15a} we would expect that a closer
inspection would produce a factor of $1/t^4$). For $v \rightarrow 0$
we have $\la_1^- = 0$ and $\la_1^+ = \p/2$, which when inserted
into (\ref{i2r1}) yields an explicit result for the leading
large-$t$ asymptotics of the dynamical part of the auto-correlation
function
\begin{equation}
     I_2 (0,t) \sim \frac{\re^{\i v_{c_2} t}}{J \p t}
                 \frac{8 (q^2;q^2)^4 (-q^4;q^4)^2 (q^8;q^8,q^8)^4}
		      {(q^{- 2} - q^2) (q^4;q^4)^6 (q^4;q^8,q^8)^4}
\end{equation}
(for field theory predictions in the critical regime
see \cite{PWA08}).

Both, in R2 and R3, only one solution of the saddle point equation
is relevant. Since $f(\n, \n) = 0$ this changes the algebraic
contribution to the asymptotics of $I_2$,
\begin{equation} \label{i2r2}
     I_2 (m, t) \sim \frac{[\nabla^2 f] (\la_j, \la_j)}{4 \p t^2} \cdot
                     \frac{\re^{2t g(\la_j)}}{g''(\la_j)^2} \epp
\end{equation}
Here $j = 2, 3$. $\la_2 = \p/4 + \i y$, where $y \in [0, \g/2]$, and
$\la_3 = \i \g/2 + x$, where $x \in [\p/4, \p/2]$. Note that the
reasons for having only one relevant saddle point are different
in R2 and R3. In R2 the steepest descent contour can only pass
through one of the points, in R3 one of the points is sub-leading.
In R2 the phase $g(\la_2)$ has a negative real part and a non-vanishing
imaginary part. The asymptotics is oscillating and exponentially
decaying. In R3 $g(\la_3)$ is real negative, we face pure exponential
decay. The four-spinon contributions are sub-leading in R2 and R3,
since they would produce factors of $4 t g(\la_j)$ in the exponent.
In the static limit, $v \rightarrow \infty$, we obtain a simple
explicit expression which we derived in \cite{DGKS15a} and do
not reproduce here.

Amazingly, it is possible to obtain the saddle-point values
$g(\la)$, $g''(\la)$, $\la = \la_1^\pm, \la_2, \la_3$ and
$[\nabla^2 f] (\la_j, \la_j)$, $j = 2, 3$ as explicit algebraic
functions of $v$. This is due to the fact that (\ref{pp}) and
(\ref{disp}) allow us to rewrite the saddle-point equation as
\begin{equation} \label{sp2}
     v \e (p) - v_{c_1} v_{c_2} \cos(p) \sin (p) = 0 \epp
\end{equation}
This can be solved for $z = \cos^2 (p)$ at the saddle
points. Introducing the rescaled velocity parameter $r =
v/{\sqrt{v_{c_1} v_{c_2}}}$ we obtain two solutions
\begin{equation} \label{zpm}
     2 z_\pm = 1 + r^2 \pm 
        \begin{cases}
	   \sqrt{(r_1^2 - r^2)(r_2^2 - r^2)} & \text{in R1, R3,} \\[2ex]
	   \i \sqrt{(r^2 - r_1^2)(r_2^2 - r^2)}
	    & \text{in R2.}
        \end{cases}
\end{equation}
Here $r_1^2 = v_{c_1}/v_{c_2}$ and $r_2^2 = v_{c_2}/v_{c_1}$. Hence,
$0 < r < r_1$ in R1, $r_1 < r < r_2$ in R2 and $r_2 < r$ in R3.
From (\ref{zpm}) we obtain $\cos (p)$ and $\sin (p)$ and therefore
$\re^{\i p}$ at the saddle points. Then $\e(p)$ follows from (\ref{sp2}).

This leads us to
\begin{equation} \label{exptgsaddle}
     \re^{t g(\la_1^\pm)} = (-1)^m
        \exp \bigl\{ - \i m \bigl(\arctg \sqrt{1/z_\pm - 1}
                           - r^{-2} \sqrt{z_\pm (1 - z_\pm)} \bigr) \bigr\}
\end{equation}
in R1. In R2 we find
\begin{align}
     \re^{2t g (\la_2)} & = \bigl(r/k\bigr)^m
        \bigl( \re^{\i \ph_0/2} - \i \sqrt{k'} \re^{- \i \ph_1/2} \bigr)^{2m}
	\exp\biggl\{ \frac{2\i m \sqrt{k'}}{rk} \re^{\i (\ph_0 - \ph_1)/2} \biggr\}
	\epc \\
     \ph_s & = \arcctg \biggl(
               \frac{1 + (-1)^s r^2}
	            {\sqrt{(r^2 - r_1^2)(r_2^2 - r^2)}} \biggr) \epp
\end{align}
Finally, in R3,
\begin{equation}
     \re^{2t g(\la_3)} =
        \exp \Bigl\{ 2m \bigl(\ln (\sqrt{z_-} - \sqrt{z_- - 1})
                           + r^{-2} \sqrt{z_- (z_- - 1)} \bigr) \Bigr\} \epp
\end{equation}
For $g''(\la)$ we obtain
\begin{equation}
     \frac{\p^2 g''(\la)}{v (2 k K)^2}
          = \frac{z_+ - z_-}{r^2} \times
	    \begin{cases}
	       \pm \i \sqrt{z_\pm (1 - z_\pm)} & \text{$\la = \la_1^\pm$} \\[1ex]
	       \frac{\i r \sqrt{k'}}{k} \re^{\i (\ph_0 - \ph_1)/2}
	       & \text{$\la = \la_2$} \\[1ex]
	       - \sqrt{z_- (z_- - 1)} & \text{$\la = \la_3$.}
	    \end{cases}
\end{equation}
It is not hard to see that the only zeros of this function as
a function of $v$ are $v_{c_1}$ and $v_{c_2}$. Thus, except
at these points, which mark the transition between different
asymptotics regimes, the saddle points are of first order.
Exactly at the transition the saddle points are of second
order implying that the algebraic part decays as $t^{- 4/3}$.
For the saddle-point values of $[\nabla^2 f] (\la, \la)$ in
R2 and R3 we obtain
\begin{multline} \label{gradfsaddle}
     [\nabla^2 f] (\la, \la) \\ =
        \frac{512 (q^2;q^2)^4 (q^4;q^4)^2 (q^4;q^4,q^4)^8}
             {(q^2;q^4)^2 (q^2;q^4,q^4)^8}
	     \biggl[ \frac{z (-1)^m (q;q^2)^4}{(1 - q^{-1})^2} +
	     \frac{(1 - z) (-q;q^2)^4}{(1 + q^{-1})^2} \biggr] \epc
\end{multline}
where $z = z_+$ if $\la = \la_2$ in R2 and $z= z_-$ if $\la = \la_3$
in R3.
\section{Discussion}
\begin{figure}
\begin{center}
\includegraphics[width=.70\textwidth]{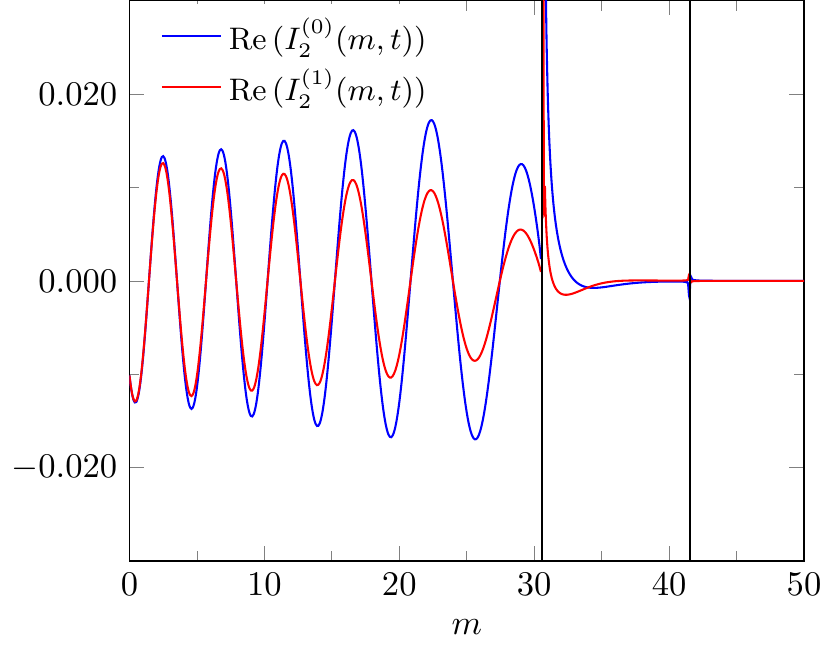}
\caption{\label{fig:fixed_t_splitRe} Real part of the two
incommensurate contributions to $I_2 (m,t)$ as a function
of $m$ for fixed $t = 4$ and $\D = 2.375$ ($I_2^{(0)}$ blue,
$I_2^{(1)}$ red). Vertical lines separate different asymptotic
regimes, first line $m = v_{c_1} t$, second line $m = v_{c_2} t$.}
\end{center}
\end{figure}

\begin{figure}
\begin{center}
\includegraphics[width=.70\textwidth]{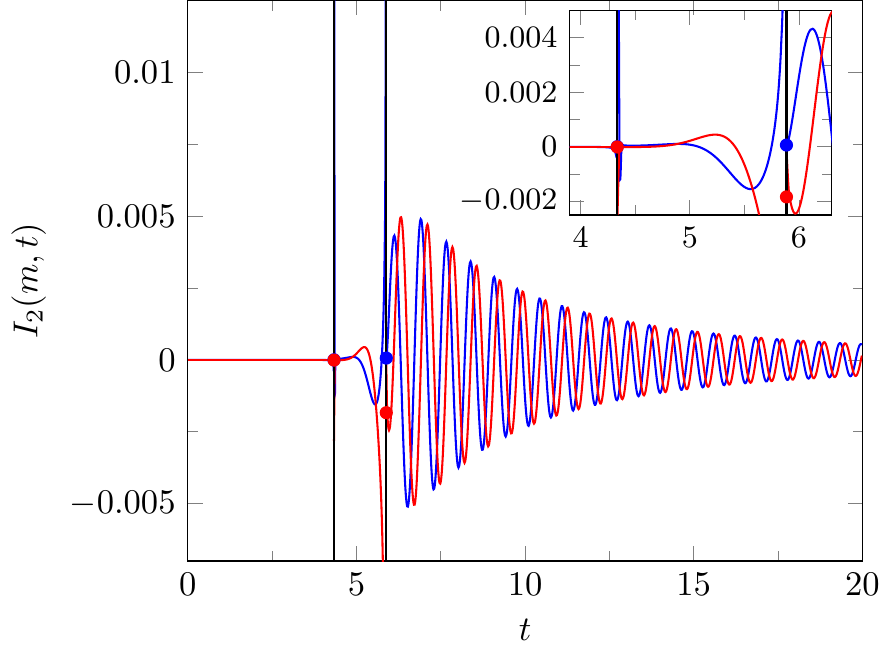}
\caption{\label{fig:fixed_m_With_Inset} $I_2 (m,t)$ as a function
of $t$ for fixed $m = 45$ and $\D = 2.375$ (real part blue,
imaginary part red). Vertical lines separate different asymptotic
regimes, first line $t = m/v_{c_2}$, second line $t = m/v_{c_1}$.
Dots denote asymptotics values exactly at the boundaries between
the different regimes. From the inset one can imagine the true
behaviour of the integral in the vicinity of the boundaries.}
\end{center}
\end{figure}

Using the above explicit expressions (\ref{ampfun}), (\ref{spnu}),
(\ref{i2r1}), (\ref{i2r2}) and (\ref{exptgsaddle})-(\ref{gradfsaddle})
we can easily plot our result.
\begin{figure}
\begin{center}
\includegraphics[width=.75\textwidth]{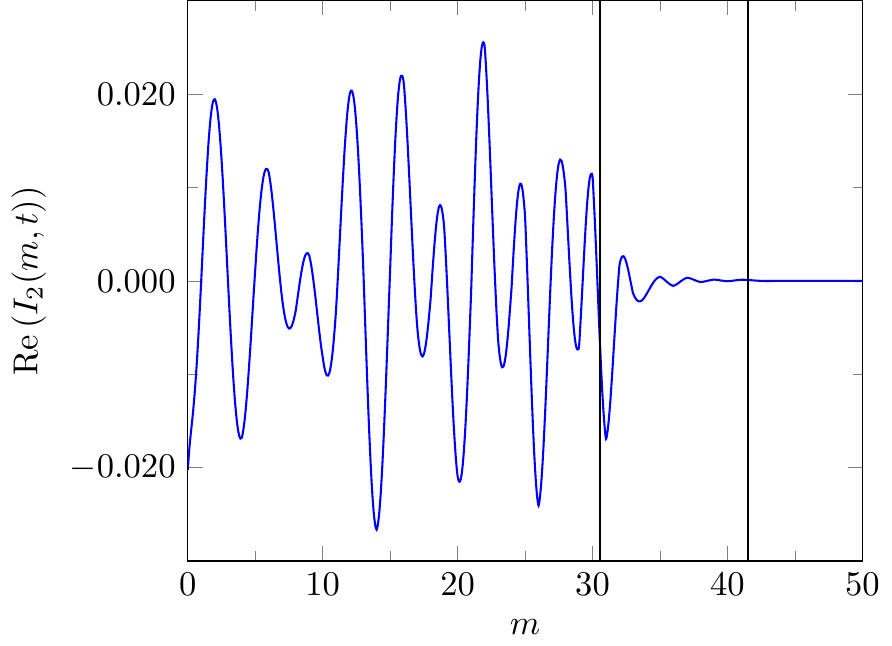}
\caption{\label{fig:fixed_t_fullRe} Real part of $I_2 (m,t)$ as a function
of $m$ for fixed $t = 4$ and $\D = 2.375$. Data points calculated for
$m \in {\mathbb Z}$ and connected by means of splines. Vertical lines
separate different asymptotic regimes, first line $m = v_{c_1} t$,
second line $m = v_{c_2} t$.}
\end{center}
\end{figure}
Figure~\ref{fig:fixed_t_fullRe} shows the asymptotics of $I_2$,
representing the leading dynamical contribution to the correlation
function $\<\s_1^z \s_{m+1}^z (t)\>$ (see (\ref{twospexp})),
for a fixed time $t$ as a function of the distance~$m$.
We see the typical features of wave propagation in a dispersive
medium as they are familiar from electrodynamics \cite{Sommerfeld14,%
Brillouin14}. The wave excited locally at $m = 1$ and $t = 0$
contains all frequency components and hence spreads out with
the maximal possible group velocity $v_{c_1}$. In fact it is
easy to see from (\ref{disp}) that
\begin{equation}
     \max_{p \in [- \p/2, \p/2]} |\e' (p)| = v_{c_1}
\end{equation}
(which is also equal to the band width $\e (0) - \e (\p/2)$).
In non-relativistic spin systems with local interactions such
a maximal group velocity always exists due to the existence
of a Lieb-Robinson bound \cite{LiRo72,CheneauBarmettler}.
As in signal processing in dispersive media the wave front
at $v_{c_1} t$ is preceded by a precursor \cite{Sommerfeld14,%
Brillouin14} extending from $m = v_{c_1} t$ to $m = v_{c_2} t$
and decaying in forward direction. In terms of `band parameters'
$v_{c_2}$ can be interpreted as twice the band centre $v_{c_2} =
- (\e (0) + \e (\p/2))$. The irregular appearance of the
wave train in Figure~\ref{fig:fixed_t_fullRe} is due to
the interference of commensurate and incommensurate
components. In fact, if we split $I_2$ as $I_2 (m, t) =
I_2^{(0)} (m, t) (-1)^m + I_2^{(1)} (m, t)$, the two contributions
$I_2^{(0)}$ and $I_2^{(1)}$ look more regular (see
Figure~\ref{fig:fixed_t_splitRe}). Figure~\ref{fig:fixed_m_With_Inset}
shows the same wave train as observed at a fixed site as a function of time.

The appearance of the two different time-like and space-like
regimes R1 and R3 is a consequence of the existence of the
Lieb-Robinson bound. But the existence of the precursor regime
R2 is not in accordance with common intuition which expects
a simpler light-cone picture as in relativistic quantum field
theories. This may be one of the reasons why the sine-Gordon
model has often been used as an effective model for the XXZ
chain in the massive antiferromagnetic regime.


Our analysis allows us two interesting conclusions. First: since
the analytic behaviour of the correlation functions changes
across the Stokes lines $v = v_{c_1}$ and $v = v_{c_2}$ the
naive analytic continuation of an effective field theory, describing
the static correlations at large distances, to the time axis
would predict the wrong large-time asymptotics. Three separate
theories for regimes R1, R2, R3 may be needed. Second: a
general theory predicting the behaviour of auto-correlation
functions is the theory of spin diffusion (see e.g.\ \cite{FaMc98}).
It predicts a decay of correlation functions $\sim 1/\sqrt{t}$.
Hence, naively the $1/t$ behaviour of the Heisenberg-Ising chain
would be called non-diffusive. However, in our saddle-point
integration we have obtained a factor of $1/\sqrt{t}$ per integral
which may be interpreted as a factor of $1/\sqrt{t}$ per spinon,
or diffusion of spinons. The $1/t$ behaviour would then be
attributed to the fact that spinons can only be created in pairs.

To avoid misunderstandings we wish to point out that there
is no obvious relation of our result for the exact asymptotics of
the two-point function to the result of \cite{BKM98}, where
the two-spinon contribution to the dynamical structure factor
(the Fourier transform of the dynamical two-point function) was
calculated. First of all \cite{BKM98} deals with the transversal
case, while we are treating the longitudinal case. Second,
to our knowledge there is no simple way to obtain the long-time,
large-distance asymptotics in real space-time by Fourier techniques.

\section{Summary}
We have obtained asymptotically exact and explicit results for
the long-time large-distance asymptotics of the longitudinal
dynamical two-point functions of the Heisenberg-Ising chain in
the easy-axis regime. Instead of a summary let us list the
main features that might be observable in experiments:
\begin{enumerate}
\item
Two critical velocities $v_{c_1}$ and $v_{c_2}$ and a precursor.
\item
A superposition of commensurate and incommensurate components.
\item
A $1/t$ decay and an oscillation with frequency $v_{c_2}$ in
the auto-correlation function.
\end{enumerate}
\noindent {\bf Acknowledgment.}
The authors acknowledge helpful discussions with K.~Fabricius,
M.~Karbach, A.~Kl\"umper, R.~Konik, J.~Sirker and R.~Weston.
MD and FG acknowledge financial support by the Volkswagen
Foundation and by the DFG under grant number Go 825/7-1. KKK
is supported by the CNRS. His work has been partly financed
by a Burgundy region PARI 2013-2014 FABER grant `Structures
et asymptotiques d'int\'egrales multiples'. KKK also enjoys
support from the ANR `DIADEMS' SIMI 1 2010-BLAN-0120-02. JS is
supported by a JSPS Grant-in-Aid for Scientific Research (C)
No.\ 15K05208.


\providecommand{\bysame}{\leavevmode\hbox to3em{\hrulefill}\thinspace}
\providecommand{\MR}{\relax\ifhmode\unskip\space\fi MR }
\providecommand{\MRhref}[2]{%
  \href{http://www.ams.org/mathscinet-getitem?mr=#1}{#2}
}
\providecommand{\href}[2]{#2}

\end{document}